\documentclass[a4paper,11pt]{article}

\usepackage{amsmath,amssymb,array,amsthm,amsxtra,overpic,bbm,bbold,bm,epsfig,multirow,longtable,booktabs}
\usepackage{url}
\usepackage {indentfirst} 
\usepackage[utf8]{inputenc}
\usepackage[margin=1in]{geometry}
\usepackage[T1]{fontenc}
\usepackage{booktabs,multirow}
\usepackage{bbold}
\usepackage{subcaption}
\usepackage{amsmath}
\usepackage{mathtools,amssymb,braket,dsfont}
\usepackage{graphicx}
\usepackage{multicol}
\usepackage{multirow}
\usepackage{float}
\usepackage{cancel}
\usepackage[normalem]{ulem}
\usepackage[dvipsnames]{xcolor}
\usepackage[final]{hyperref}
\hypersetup{
	colorlinks=true,
	linkcolor=orange,
	citecolor=blue,
	filecolor=magenta,
	urlcolor=blue
}
\usepackage[numbers, sort&compress]{natbib}
\usepackage[utf8]{inputenc}
\usepackage[T1]{fontenc}
\usepackage{comment}

\begin{document}
\begin{titlepage}
\begin{center}

{\LARGE \bf \boldmath Universal two-zero texture in SO(10): 
implications of JUNO and realization from non-invertible symmetries} \\[1cm]

Zi-Qiang~Chen,$^{a,b,c,}$\footnote{E-mail: \texttt{chenziqiang22@mails.ucas.ac.cn}}
Wen-Hao~Jiang,$^{a,b,c,}$\footnote{E-mail: \texttt{jiangwenhao25@mails.ucas.ac.cn}}
and
Ye-Ling~Zhou$^{a,}$\footnote{E-mail: \texttt{zhouyeling@ucas.ac.cn}}
\\[-2mm]
\end{center}
\vspace*{0.50cm}
\centerline{$^{a}$\it School of Fundamental Physics and Mathematical Sciences,}
\centerline{\it Hangzhou Institute for Advanced
Study, UCAS, Hangzhou 310024, China}
\centerline{$^{b}$\it Institute of Theoretical Physics, Chinese Academy of Sciences, Beijing 100190, China}
\centerline{$^{c}$\it University of Chinese Academy of Sciences, Beijing 100049, China}

\vspace*{1cm}

\begin{abstract}
We apply the universal two-zero texture (UTZT) to all quark and lepton mass
matrices in the SO(10) grand unified framework. 
With charged fermion masses fixed at their best-fit values, this texture contains only seven free parameters to account for nine flavor observables, rendering it highly predictive. 
Motivated by the recent JUNO indication in favor of the normal ordering of light neutrino masses, we perform an updated analysis of the UTZT in SO(10). 
The texture remains fully compatible with all current flavor data.
The Dirac phase is predicted mainly in two regions, one of which matches very well with current data. A meV-scale $m_{\beta\beta}$ is predicted, beyond the sensitivity bound of future neutrinoless double beta decay measurements.
We further explore the origin of the UTZT from non-invertible symmetries, without introducing additional low-energy degrees of freedom.
We show that the UTZT can be realized through non-invertible selection rules arising from the $Z_3$ gauging of $Z_N$, with a minimal realization corresponding to $N=7$. 
\end{abstract}

\end{titlepage}

\section{Introduction}

The texture-zero approach, after its proposal \cite{Fritzsch:1977za, Weinberg:1977hb, Wilczek:1977uh}, is remaining an enduring solution to the flavor puzzle in particle physics. 
It assumes a few vanishing entries in the fermion mass matrices, which enforce correlations between fermion masses and mixing,  and  the number of free parameters in the flavor space is efficiently reduced. 
A widely discussed example is a type of two-zero texture that a fermion mass matrix takes the form
\begin{eqnarray} \label{eq:two_zeros} 
	M_{f} &\sim& \left( \begin{matrix} 0 & \times & 0 \\
		\times  & \times & \times \\ 
		0 & \times & \times \\
	\end{matrix} \right) \,,
\end{eqnarray} 
in which ``$\times$'' denotes a non-vanishing arbitrary value. The two zeros correspond to the vanishing (1,1) and (1,3) entries in the matrix, and the $(3,1)$ entry is usually not considered to be independent based on the requirement of matrix Hermiticity or symmetry in specific models. It is extended from the three-zero Fritzsch texture \cite{Fritzsch:1977vd, Fritzsch:1979zq}, where $M_f$ is Hermitian and (2,2) entry vanishes.
A prominent textured ansatz is the four-zero texture \cite{Du:1992iy, Fritzsch:1995nx, Xing:1996hi}, wherein both the up- and down-type quark mass matrices are assumed to be Hermitian and take the above form. It takes advantages of its analytical simplicity \cite{Fritzsch:2002ga,Xing:2003yj} and solving the strong CP problem \cite{Xing:2015sva, Fritzsch:2021ipb}. 

The two-zero texture is applied to the neutrino mass matrix $M_\nu$ in the basis where charged lepton mass matrix $M_l$ is first considered as diagonal \cite{Frampton:2002yf, Xing:2002ta, Xing:2002ap}; see \cite{Fritzsch:2011qv, Denton:2023hkx, Denton:2026wpy} for discussions on how well fitting to new data. However, a leptonic version of four-zero texture, that both $M_l$ and $M_\nu$ have two-zero textures is proposed in \cite{Xing:2003zd}, which allows the generalization to universal two-zero textures (UTZT) for all lepton mass matrices, including Dirac and Majorana mass matrices for right-handed neutrinos (RHNs), in the seesaw framework.
The UTZT includes enough parameter space to fit data and has been continuously studied in light of updated data from neutrino measurements \cite{Zhou:2012ds, Barranco:2012ci, Fritzsch:2012rg, Liao:2013kix, Yang:2020qsa, Yang:2021xob}.

Motivated by its phenomenological success, the UTZT for all quark and lepton masses has been embedded into SO(10) grand unified theories (GUTs) \cite{Fang:2024qtx}. Below, we denote it as UTZT-SO(10). An important advantage of the GUT version is the substantial reduction of the dimensionality of the flavor-parameter space. Since all quarks and leptons, including RHNs, are unified into the 16-dimensional spinorial representation, nontrivial correlations among fermion masses and mixing parameters naturally emerge. With the steadily improving precision of neutrino oscillation measurements, such quark-lepton correlations have become increasingly powerful probes of GUT scenarios.
Extensive numerical analyses have shown a preference for the normal ordering (NO) for light neutrino masses in several compelling and realistic GUT scenarios \cite{Altarelli:2013aqa, Babu:2016bmy, Babu:2018qca, Ohlsson:2019sja, Mummidi:2021anm,Fu:2022lrn, Fu:2023mdu,Babu:2025wop}, although inverted ordering is not fully ruled out \cite{Dueck:2013gca,Saad:2022mzu,Babu:2024ahk}. 
This is further confirmed by a recent multi-dimensional numerical scan \cite{Chen:2025afg} taking account of JUNO's first physical results \cite{JUNO:2025gmd}. This finding is in good agreement with the latest JUNO data on the indication of NO at $2.3~\sigma$ \cite{JUNO_2026}. 

Texture zeros can be realized in the discrete flavor symmetries, either via Abelian \cite{Grimus:2004hf} or non-Abelian symmetries \cite{GonzalezCanales:2012blg}, however, both at the price of extra copies of Higgses. 
This approach is applied to the UTZT by employing a $Z_6$ symmetry in the seesaw framework \cite{Zhou:2012ds}. Applying Abelian discrete symmetries to mass textures in the GUT framework has also been studied in \cite{Ferreira:2015jpa,Ivanov:2015xss,Ohlsson:2021lro,Lindestam:2021dyk}. A UTZT-SO(10) model is constructed in \cite{Fang:2024qtx}, where the contribution of extra Higgses must be carefully treated to avoid the Landau pole in the gauge coupling renormalization-group running and to satisfy constraints from proton-decay experiments. Modular symmetries provide an alternative possibility to realize texture zeros if the theory is supersymmetric \cite{Zhang:2019ngf,Lu:2019vgm}.

Non-invertible symmetries (NIS's) were implemented in particle physics in recent years \cite{Choi:2022jqy, Cordova:2022ieu, Cordova:2022fhg, Cordova:2024ypu, Kobayashi:2024yqq}.
A bottom-up approach of applying NIS to flavor model building has been proposed in \cite{Kobayashi:2024cvp}. In particular, the nearest neighbor interaction texture \cite{Branco:1988iq}, an asymmetric extension of the Fritzsch texture, is realized in the $Z_2$ gauging of $Z_N$. The NIS provides the most economical approach to texture zeros as it requires no new particles to be introduced at low energies.
A systematic study of texture zeros in the neutrino sector via the $Z_2$ gauging of $Z_N$ in the minimal seesaw framework is performed in \cite{Jiang:2025psz}. Textures in both charged lepton and neutrino sectors are further explored in the same symmetries in \cite{Kobayashi:2025ldi}. Three-zero textures for the down-quark masses are proposed in the GUT framework based on the symmetries \cite{Kobayashi:2025thd}. 
In the charged lepton flavor basis, all two-zero textures for neutrino masses are achieved in the $Z_3$ gauging of $Z_N$ \cite{Qu:2026omn}. 

In this work, we will discuss the up-to-date fitting of the UTZT-SO(10) and its realization in NIS's. 
Compared with former works, main novelties of this paper are addressed below.
\begin{itemize}
\item The UTZT-SO(10) framework remains compatible with the new data of precision neutrino measurements.
\item We adopt the numerical strategy of Ref.~\cite{Chen:2025afg} to perform a more extensive scan of the parameter space. This scan reveals additional disconnected viable branches in the allowed ranges of \(\delta^l\), including regions close to maximal leptonic CP violation and regions of \(m_{\beta\beta}\), which were not discovered in Ref.~\cite{Fang:2024qtx}.
\item Since no new particles are introduced at low energies in the NIS framework, this approach automatically avoids the problems of gauge coupling renormalization group (RG) running and experimental constraints of proton decay. 
\end{itemize}

The rest of this paper is organized as follows. In section~\ref{sec:2}, we review the fermion masses and mixing of the UTZT-SO(10), where the quark-lepton correlation induced by GUT is emphasized. 
The numerical analysis of the fermion masses and mixing fitting to the most recent JUNO data is performed in section~\ref{sec:3}. We consider only the NO case for neutrino masses, since NO is supported by recent JUNO result and no viable points in the IO case are found in former fit \cite{Fang:2024qtx}. 
In section~\ref{sec:4}, we discuss how the UTZT is realized via non-invertible selection rules. 
An explicit model based on the $Z_3$ gauging of $Z_N$ will be presented, while realizations in other NIS's will also be discussed. 
We conclude in  section~\ref{sec:5}.

\section{Fermion masses and mixing in UTZT-SO(10)}\label{sec:2}
In this section, we briefly revisit the UTZT-SO(10), where light neutrino masses are generated via the Type-I+II seesaw mechanism after spontaneous symmetry breaking. We follow the discussion in Ref. \cite{Fang:2024qtx}, where detailed derivations of flavor mixing in two-zero textures and a parametrization of quark-lepton correlations are provided. New contents of the present work include the updated fit and the more comprehensive parameter scan in section \ref{sec:3} and the realization of the UTZT from non-invertible selection rules in section \ref{sec:4}.

We assume the following breaking chain of gauge symmetries in the non-supersymmetric framework 
\begin{eqnarray} \label{eq:breaking_chain}
SO(10) &\underset{\mathbf{54}}{\overset{M_{\mathrm{GUT}}}\longrightarrow}& G_{422}^{\rm C} = SU(4)_c \times SU(2)_L \times SU(2)_R \times Z_2^{\rm C} \nonumber\\
&\underset{\mathbf{45}}{\overset{M_2}\longrightarrow}& G_{3221} = SU(3)_c \times SU(2)_L \times SU(2)_R \times U(1)_X \nonumber\\ 
&\underset{\overline{\mathbf{126}}}{\overset{M_1}\longrightarrow}& G_{\mathrm{SM}} = SU(3)_c \times SU(2)_L \times U(1)_Y\,,
\end{eqnarray}
where $M_{\rm GUT}$ and $M_{1,2}$ denote breaking scales of GUT and intermediate gauge symmetries, respectively, with the responsible Higgs multiplet indicated in the subscript. $Z_2^{\rm C}$ is the parity symmetry that relates left-handed particles to their right-handed charge-conjugate partners. This chain has been well studied in \cite{Bertolini:2009qj,King:2021gmj} and denoted as II3 in \cite{King:2021gmj}. The CP symmetry is also imposed above the GUT scale, essential for the achievement of Hermitian Yukawa matrices at low energies \cite{Fang:2024qtx}. Assuming the existence of the UTZT at the GUT scale, we then discuss its analytical properties and reparameterize the fermion Yukawa matrices.

\subsection{Quark-lepton correlation in SO(10) GUTs}
In SO(10) GUTs, each generation of SM fermions, including RHNs, is embedded in a single $\mathbf{16}$-dimensional spinor representation. The Higgs sector contains $\mathbf{10}_H$, $\overline{\mathbf{126}}_H$, $\mathbf{120}_H$, $\mathbf{54}_H$, and $\mathbf{45}_H$. The multiplets $\mathbf{10}_H$, $\overline{\mathbf{126}}_H$, and $\mathbf{120}_H$ couple to the fermion bilinear $\mathbf{16}\times\mathbf{16}$ and generate SO(10)-invariant Yukawa interactions. The $\mathbf{45}_H$ and $\mathbf{54}_H$ fields are introduced for gauge symmetry breaking as presented in Eq.~\eqref{eq:breaking_chain}, while $\overline{\mathbf{126}}_H$ also breaks $B-L$ and gives Majorana masses to RHNs. The Yukawa couplings of fermions are given by
\begin{eqnarray} \label{eq:Yukawa_couplings_SO(10)}
    -\mathcal{L}_{\mathrm{Y}} = (A)_{\alpha\beta} \; \mathbf{16}^\alpha_F \mathbf{16}^\beta_F \mathbf{10}_H \; + \; 
    (B)_{\alpha\beta} \; \mathbf{16}^\alpha_F \mathbf{16}^\beta_F \overline{\mathbf{126}}_H \; + \; 
    (C)_{\alpha\beta} \; \mathbf{16}^\alpha_F \mathbf{16}^\beta_F \mathbf{120}_H \; + \mathrm{h.c.},
\end{eqnarray}
assuming $\mathbf{10}_H$, $\overline{\mathbf{126}}_H$, and $\mathbf{120}_H$ are all complex fields. There are additional terms 
\begin{eqnarray} \label{eq:Yukawa_couplings_add}
    (A')_{\alpha\beta} \; \mathbf{16}^\alpha_F \mathbf{16}^\beta_F \mathbf{10}_H^* \; + \; 
    (C')_{\alpha\beta} \; \mathbf{16}^\alpha_F \mathbf{16}^\beta_F \mathbf{120}_H^* \; + \mathrm{h.c.},
\end{eqnarray}
which can be forbidden by imposing a Peccei-Quinn (PQ) $U(1)$ symmetry.

The decomposition of $\mathbf{16} \times \mathbf{16}$ into $\mathbf{10}_S$, $\mathbf{126}_S$ and $\mathbf{120}_A$ indicates that the Yukawa coupling matrices generated by $\mathbf{10}_H$ and $\overline{\mathbf{126}}_H$, i.e., $(A)_{\alpha\beta}$ and $(B)_{\alpha\beta}$, are symmetric, while $(C)_{\alpha\beta}$ is antisymmetric. The structure of Yukawa couplings can be further constrained by additional symmetries. We impose the CP symmetry above the GUT breaking scale, leading to real coupling matrices $A$, $B$ and $C$ \cite{Grimus:2006rk}. 
In order to obtain CP violation in the mixing matrix, we consider spontaneous CP violation as originally proposed in \cite{Dutta:2004hp}. It is achieved by assuming real VEVs of ${\bf 10}_H$ and $\overline{\bf 126}_H$ and a purly imaginary VEV of ${\bf 120}_H$. 

The resulting Yukawa couplings of up-type quarks, down-type quarks, charged leptons and neutrinos are expressed as \cite{Dutta:2004zh,Dutta:2005ni, Altarelli:2010at, Dutta:2009ij}:
\begin{eqnarray} \label{eq:Yukawa}
	Y_{u} &=& H + r_2 F + i r_3 G \,, \nonumber\\
	Y_{d} &=& r_1 (H+ F+ i G) \,, \nonumber\\
	Y_\nu &=& H - 3 r_2 F + i c_\nu G  \,, \nonumber\\
	Y_{e} &=& r_1 (H - 3 F + i c_{e} G) \,, 
\end{eqnarray}
where $H$, $F$ and $G$ are $3\times 3$ real matrices proportional to the coupling matrices $A$, $B$ and $C$ respectively, and the parameters $r_1$, $r_2$, $r_3$, $c_\nu$ and $c_{e}$ are all real free parameters parametrizing the Higgs VEV contributions. Consequently, the Yukawa coupling matrices remain Hermitian. The light neutrino mass matrix is generated by the type-(I+II) seesaw mechanism, given by
\begin{eqnarray} \label{eq:nu_mass}
	M_\nu &=& m_L \, F - m_R \, Y_\nu F^{-1} Y_\nu^T \,,
\end{eqnarray}
where $m_L$ and $m_R$ are generated by VEVs of triplet Higgses of $SU(2)_L$ and $SU(2)_R$, respectively, both decomposed from $\overline{\bf 126}_H$ of SO(10). The RHN mass matrix is given by
\begin{eqnarray} \label{eq:N_mass}
	M_R = \frac{v^2}{2 m_R} F \,,
\end{eqnarray}
where $v$ is the VEV of the SM Higgs doublet.

\subsection{Fermion masses and mixing with UTZT-SO(10)}
A characteristic feature of GUT theory is that all fermion mass matrices are correlated with each other, which efficiently reduces the number of free parameters in the model. The UTZT for all fermion mass matrices can further reduce the number of free parameters. The UTZT-SO(10) has been shown to be compatible with current experimental data, while providing predictions for neutrino Majorana masses that can be tested in future experiments. To illustrate the correlations between fermion masses and mixing, Yukawa matrices of charged leptons and neutrinos, as well as the parameterized matrix $F$, are expressed in terms of those of quarks as
\begin{eqnarray} \label{eq:Ye}
	Y_{e} &=& -\frac{4\, r_1}{r_2-1} \operatorname{Re} Y_{u} + \frac{r_2+3}{r_2-1} \operatorname{Re} Y_{d} + i\, c_{e} \operatorname{Im} Y_{d} \,, \nonumber\\
	Y_\nu &=& - \frac{3r_2+1}{r_2-1} \operatorname{Re} Y_{u} + \frac{4r_2}{r_1(r_2-1)}\operatorname{Re} Y_{d} + i \frac{c_\nu}{r_1} \operatorname{Im} Y_{d} \,, \nonumber\\
	F &=&\frac{\operatorname{Re} Y_{u}}{r_2-1}- \frac{\operatorname{Re} Y_{d}}{r_1(r_2-1)} \,.
\end{eqnarray}
From Eqs.~\eqref{eq:nu_mass} and \eqref{eq:N_mass}, the light neutrino mass matrix $M_\nu$ and the RHN mass matrix $M_R$ also depend on $Y_{u}$, $Y_{d}$, $r_1$, $r_2$ and $c_\nu$. Assuming that the Yukawa couplings matrices $A$, $B$ and $C$ have the two-zero texture, it is straightforward to show that all Yukawa coupling matrices $Y_{f}$, as well as $M_\nu$, have the same two-zero texture, i.e., the UTZT is realized in this model.

To reduce the number of free parameters, the Yukawa coupling matrices are assumed to be Hermitian, i.e.,
\begin{eqnarray}
	Y_{f}= \zeta_{f} \left(
	\begin{array}{ccc}
		0 & C_{f} & 0 \\
		C_{f}^* & \tilde{B}_{f} & B_{f} \\
		0 & B_{f}^* & A_{f} \\
	\end{array}
	\right) \,,
	\label{eq:zero}
\end{eqnarray}
where $f=u,d,\nu,e$, the entries $A_{f}$ and $\tilde{B}_{f}$ are real, while $B_{f}$ and $C_{f}$ are complex, $A_{f}>0$ and $\zeta_{f} = \pm1$. $Y_{u}$ can be reparameterized in terms of the up-type quark Yukawa eigenvalues $y_{u}$, $y_{c}$ and $y_{t}$. $\zeta_{u}$ is set to $+1$ by a phase redefinition of $\mathbf{16}^\alpha_F$. The Hermitian Yukawa matrix $Y_{u}$ is transformed into a real and symmetric matrix by a phase rotation $P_{u} = {\rm diag}(1,\, e^{i\phi_{u}},\, e^{i\phi_{u}'})$, i.e., 
\begin{eqnarray}
	\overline{Y}_{u} \equiv P_{u} Y_{u} P_{u}^* =  \left(
	\begin{array}{ccc}
		0 & |C_{u}| & 0 \\
		|C_{u}| & \tilde{B}_{u} & |B_{u}| \\
		0 & |B_{u}| & A_{u} \\
	\end{array}
	\right) \,,
\end{eqnarray} 
where $A_{u}>0$ and $\tilde{B}_{u}$ can be either positive or negative. $\overline{Y}_{u}$ can be diagonalized by a real orthogonal matrix $O_{u}$,
\begin{eqnarray}
	\overline{Y}_{u} = O_{u} \hat{Y}_{u} O_{u}^T 
\end{eqnarray}
with $\hat{Y}_{u} = {\rm diag}(-\eta_{u} y_{u},\, \eta_{u} y_{c},\, y_{t})$, where $\eta_{u} = \pm 1$ is an undetermined sign. 
We claim that the (3,3) entry of the texture is always smaller than the largest eigenvalue, i.e., $A_u < y_t$. This is proven below. Starting from the identity,
\[
\det(Y_u-A_u I)=A_u|B_u|^2
=\det(\hat{Y}_u-A_u I)
=(-\eta_u y_u-A_u)(\eta_u y_c-A_u)(y_t-A_u)>0, 
\]
a direct analysis gives \(y_c<A_u<y_t\) for \(\eta_u=+1\), and \(y_u<A_u<y_t\) for \(\eta_u=-1\).
$O_{u}$ can be explicitly expressed in terms of $A_{u}$ and $y_{u}$, $y_{c}$, $y_{t}$ as \cite{Fritzsch:2002ga,Fritzsch:2021ipb}
\begin{eqnarray}\label{eq:Ou}
	O_{u}= 
	\begin{pmatrix}
		\sqrt{\frac{y_{c} y_{t} (A_{u}+\eta_{u}  y_{u})}{A_{u} (y_{u}+y_{c}) (\eta_{u}  y_{u}+y_{t})}} & 
		\eta_{u}  \sqrt{\frac{y_{u} y_{t} (A_{u}-\eta_{u}  y_{c})}{A_{u} (y_{u}+y_{c}) (y_{t}-\eta_{u}  y_{c})}} & 
		\sqrt{\frac{y_{u} y_{c} (y_{t}-A_{u})}{A_{u} (\eta_{u}  y_{u}+y_{t}) (y_{t}-\eta_{u}  y_{c})}} \\[5mm]
		-\eta_{u}  \sqrt{\frac{y_{u} (A_{u}+\eta_{u}  y_{u})}{(y_{u}+y_{c}) (\eta_{u}  y_{u}+y_{t})}} & 
		\sqrt{\frac{y_{c} (A_{u}-\eta_{u}  y_{c})}{(y_{u}+y_{c}) (y_{t}-\eta_{u}  y_{c})}} & 
		\sqrt{\frac{y_{t} (y_{t}-A_{u})}{(\eta_{u}  y_{u}+y_{t}) (y_{t}-\eta_{u}  y_{c})}} \\[5mm]
		\eta_{u}  \sqrt{\frac{y_{u} (y_{t}-A_{u}) (A_{u}-\eta_{u}  y_{c})}{A_{u} (y_{u}+y_{c}) (\eta_{u}  y_{u}+y_{t})}} & 
		-\sqrt{\frac{y_{c} (y_{t}-A_{u}) (A_{u}+\eta_{u}  y_{u})}{A_{u} (y_{u}+y_{c}) (y_{t}-\eta_{u}  y_{c})}} & 
		\sqrt{\frac{y_{t} (A_{u}+\eta_{u}  y_{u}) (A_{u}-\eta_{u}  y_{c})}{A_{u} (\eta_{u}  y_{u}+y_{t}) (y_{t}-\eta_{u}  y_{c})}}
	\end{pmatrix} \,.
\end{eqnarray}
Then, we get the relations
\begin{eqnarray} \label{eq:ABC_u}
	\tilde{B}_{u} &=& -\eta_{u} y_{u} + \eta_{u} y_{c} +y_{t} - A_{u} \,,\nonumber\\
	|B_{u}| &=& \sqrt{(A_{u} + \eta_{u} y_{u})(A_{u} - \eta_{u} y_{c}) (y_{t} - A_{u})/A_{u}} \,, \nonumber\\
	|C_{u}| &=& \sqrt{y_{u} y_{c} y_{t} / A_{u}} \,.
\end{eqnarray}
The reparameterization of $Y_{d}$ and $Y_{e}$ follows the same procedure as that of $Y_{u}$, except for the sign ambiguity of $\zeta_{d}$ and $\zeta_{e}$, i.e.,
\begin{eqnarray}
	Y_{d} &=& \zeta_{d} P_{d}^* O_{d} \hat{Y}_{d} O_{d}^T P_{d} \,,\nonumber\\
	Y_{e} &=& \zeta_{e} P_{e}^* O_{e} \hat{Y}_{e} O_{e}^T P_{e} \,,
\end{eqnarray}
where $\{\hat{Y}_{d},\,O_{d},\,P_{d}\}$ and $\{\hat{Y}_{e},\,O_{e},\,P_{e}\}$ have the same structure as $\{\hat{Y}_{u},\,O_{u},\,P_{u}\}$, under the substitution of $u \to d,\,c \to s,\,t \to b$ and $u \to e,\,c \to \mu,\,t \to \tau$ respectively. The CKM matrix is represented as
\begin{eqnarray} \label{CKM}
	V_{\rm CKM} = O_{u}^T P_{q} O_{d} \,,
\end{eqnarray}
where $P_{q}  = P_{u} P_{d}^* = {\rm diag}(1,\, e^{i \phi_1},\, e^{i \phi_2})$ with
$\phi_1 = \phi_{u} - \phi_{d}$ and $\phi_2 = \phi_{u}^\prime - \phi_{d}^\prime$. We refer to \cite{Fang:2024qtx} for an analytical approximation of the mixing angles in terms of quark masses, $A_{u}$, $A_{d}$, phases $\phi_{1,2}$ and signs $\eta_{u,d}$, which will not be repeated here.

In brief, there are four free parameters in the quark sector, $\{A_{u},\,A_{d},\,\phi_1,\,\phi_2\}$, with the ambiguity of signs $\{\zeta_{d},\,\eta_{u},\,\eta_{d}\}$. 

In the lepton sector, the first relation of Eq.~\eqref{eq:Ye} provides four constraints on $Y_{e}$. Combining these with
\begin{eqnarray} \label{eq:ABC_e}
	\tilde{B}_{e} &=& -\eta_{e} y_{e} + \eta_{e} y_\mu +y_\tau - A_{e} \,,\nonumber\\
	|B_{e}| &=& \sqrt{(A_{e} + \eta_{e} y_{e})(A_{e} - \eta_{e} y_\mu) (y_\tau - A_{e})/A_{e}} \,, \nonumber\\
	|C_{e}| &=& \sqrt{y_{e} y_\mu y_\tau / A_{e}} \,,
\end{eqnarray}
we can solve for four parameters, $A_{e},\,c_{e},\,r_1,\,r_2$, as well as the sign $\zeta_e$, given $\eta_e$, eigenvalues of the charged lepton Yukawas and parameters in the quark sector. The remaining free parameters in the lepton sector are $c_\nu$, $m_L$ and $m_R$, which will be further used to determine $M_\nu$. The sign $\zeta_\nu$ cancels out in Eq.~\eqref{eq:nu_mass} and thus does not represent a free parameter.

Given $F$ and $Y_\nu$ follow the two-zero textures, it has been proven that $M_\nu$ via Type-I+II seesaw formula also takes the two-zero texture \cite{Xing:2003zd}, however not Hermitian but complex symmetric. Phases of non-vanishing entries of $M_\nu$ cannot be fully rotated away via phase rotation. In this case, one has to diagonalize this matrix numerically. Suppose $M_\nu$ has been diagonalized by a unitary matrix $U_\nu$, i.e., $M_\nu = U_\nu \, {\rm diag}\{ m_1, m_2, m_3\} U_\nu^T$. Here, both $M_\nu$ and $U_\nu$ are determined by $\{m_L,m_R,c_\nu\}$. The PMNS matrix is then given by 
\begin{eqnarray} \label{PMNS}
	U_{\rm PMNS} = O_{e}^T P_{e} U_{\nu} \,.
\end{eqnarray}

To end this section, we summarize that, apart from the fermion Yukawa eigenvalues, there are seven free parameters in the model 
\begin{eqnarray} 
	\{A_{u},\,A_{d},\,\phi_1,\,\phi_2,\,m_L,\,m_R,\,c_\nu\} \,,
\end{eqnarray}   
with the sign ambiguities $\{\,\eta_{u},\,\eta_{d},\,\eta_{e},\zeta_{d}\}$.

\section{Parameter fitting}\label{sec:3}
Assuming the UTZT at the GUT scale, we perform a numerical analysis of fermion masses and flavor mixing, taking into account recent JUNO data. To reduce the dimensionality of the parameter space, we fix the eigenvalues of the Hermitian fermion Yukawa matrices to their best-fit values at
$M_{\rm GUT}=2\times 10^{16}~{\rm GeV}$, as provided in Ref.~\cite{Babu:2016bmy}:
\begin{align}
	y_{u}^{\text{bf}} &= 2.54 \times 10^{-6}, &
	y_{c}^{\text{bf}} &= 1.37 \times 10^{-3}, &
	y_{t}^{\text{bf}} &= 0.428,\nonumber \\
	y_{d}^{\text{bf}} &= 6.56 \times 10^{-6}, &
	y_{s}^{\text{bf}} &= 1.24 \times 10^{-4}, &
	y_{b}^{\text{bf}} &= 5.7 \times 10^{-3},\nonumber \\
	y_{e}^{\text{bf}} &= 2.70341 \times 10^{-6}, &
	y_\mu^{\text{bf}} &= 5.70705 \times 10^{-4}, &
	y_\tau^{\text{bf}} &= 9.7020 \times 10^{-3}.
\end{align}
The best-fit values and $1\sigma$ uncertainties of the CKM mixing angles and
the quark CP-violating phase are
\begin{align}
	&\theta_{12}^q = 13.028^\circ \pm 0.034^\circ, \qquad
	\theta_{23}^q = 2.783^\circ \pm 0.034^\circ, \nonumber\\ 
	&\theta_{13}^q = 0.2408^\circ \pm 0.0074^\circ, \qquad
	\delta^q = 69.16^\circ \pm 3.09^\circ.
\end{align}
In the neutrino sector, as indicated by JUNO's most recent result \cite{JUNO_2026}, we will consider only NO for light neutrino masses and use the following three latest inputs,
\begin{align}\label{eq:juno_data}
    &\Delta m^2_{21} = (7.388 \pm 0.078) \times 10^{-5}~{\rm eV}^2\,,\quad  \Delta m^2_{31} = (2.509^{+ 0.027}_{-0.025}) \times 10^{-3}~{\rm eV}^2\,, \nonumber\\
    &\sin^2\theta^l_{12} = 0.3036 \pm 0.0064\,.
\end{align}
Inputs for remaining oscillation parameters are taken from NuFIT 6.1 in NO for the IC23 fit without SK atmospheric data \cite{Esteban:2024eli},
\begin{align}\label{eq:NUFIT}
    \sin^2\theta^l_{23} &= 0.470 ^{+ 0.017}_{- 0.014}\,, \qquad
    \sin^2\theta^l_{13} = 0.02249 \pm 0.00057\,. 
\end{align}
The numerical fit is performed in two stages. For both quark and lepton sectors, we define
\begin{align}
    \chi^2_{\alpha}=\sum_{i=1}^{n_{\rm obs}}
    \left(\frac{\mathcal{O}_i-\mathcal{O}^{\rm bf}_i}
    {\sigma_{\mathcal{O}_i}}\right)^2 ,
    \qquad \alpha=q,l ,
    \label{eq:chi2}
\end{align}
where $\mathcal{O}_{i}$, $\mathcal{O}^{\rm bf}_{i}$ and $\sigma_{\mathcal{O}_i}$ denote the theoretical prediction, experimental best-fit value and $1\sigma$ uncertainty of the $i$-th observable, respectively. The minimization of $\chi^2_{\alpha}$ function is carried out with a differential evolution algorithm, following the setup tested in Ref.~\cite{Chen:2025afg}. The fitting procedure is summarized as follows. 
\begin{enumerate}
\item In the quark sector, the eigenvalues of $Y_{u}$ and $Y_{d}$ are fixed. For each point in the parameter space of Tab.~\ref{tab:Para} , the values of $A_{u}$ and $A_{d}$ ($A_{u}< y_t$ and $A_{d}<y_b$ as proven in the last section) determine $O_{u}$ and $O_{d}$ through Eq.~\eqref{eq:Ou} up to the signs $\eta_{u},\eta_{d}=\pm1$. Together with the phase differences $\phi_1=\phi_{u}-\phi_{d}$ and $\phi_2=\phi'_{u}-\phi'_{d}$, the CKM matrix is obtained from Eq.~\eqref{CKM}. For each choice of $\eta_{u},\eta_{d}=\pm1$,  the four parameters $\{A_{u},A_{d},\phi_1,\phi_2\}$ are varied to fit $\{\theta^q_{12},\theta^q_{13},\theta^q_{23},\delta^q\}$. Points satisfying $\chi_q^2<10$ are retained for the subsequent lepton-sector fit.

\item 
In the lepton sector, as presented in the previous section, $A_e$, $r_1$, $r_2$ and $c_{e}$ are determined via Eqs.~\eqref{eq:ABC_e} and ~\eqref{eq:Ye} after signs $\eta_e$, $\zeta_d$, eigenvalues of the charged lepton Yukawas, and parameters in quark sector are given. Once these parameters are fixed, $Y_\nu$ is only determined by $c_\nu$ via the second equation of Eq.~\eqref{eq:Ye}. Consequently, $M_\nu$ is determined by three parameters $\{m_L,m_R,c_\nu\}$ via Type-I+II seesaw formula in Eq.~\eqref{eq:Yukawa}. We vary these parameters  and minimize $\chi_l^2$ by  fitting $\{\theta^l_{12},\theta^l_{13},\theta^l_{23}, \Delta m^2_{21},\Delta m^2_{31}\}$.

\item The final results are selected by requiring $\chi^2_{\rm tot}=\chi_q^2+\chi_l^2<10$.
\end{enumerate}

\begin{figure}[htbp] 
	\begin{center}     		
		\includegraphics[width=0.85\textwidth]{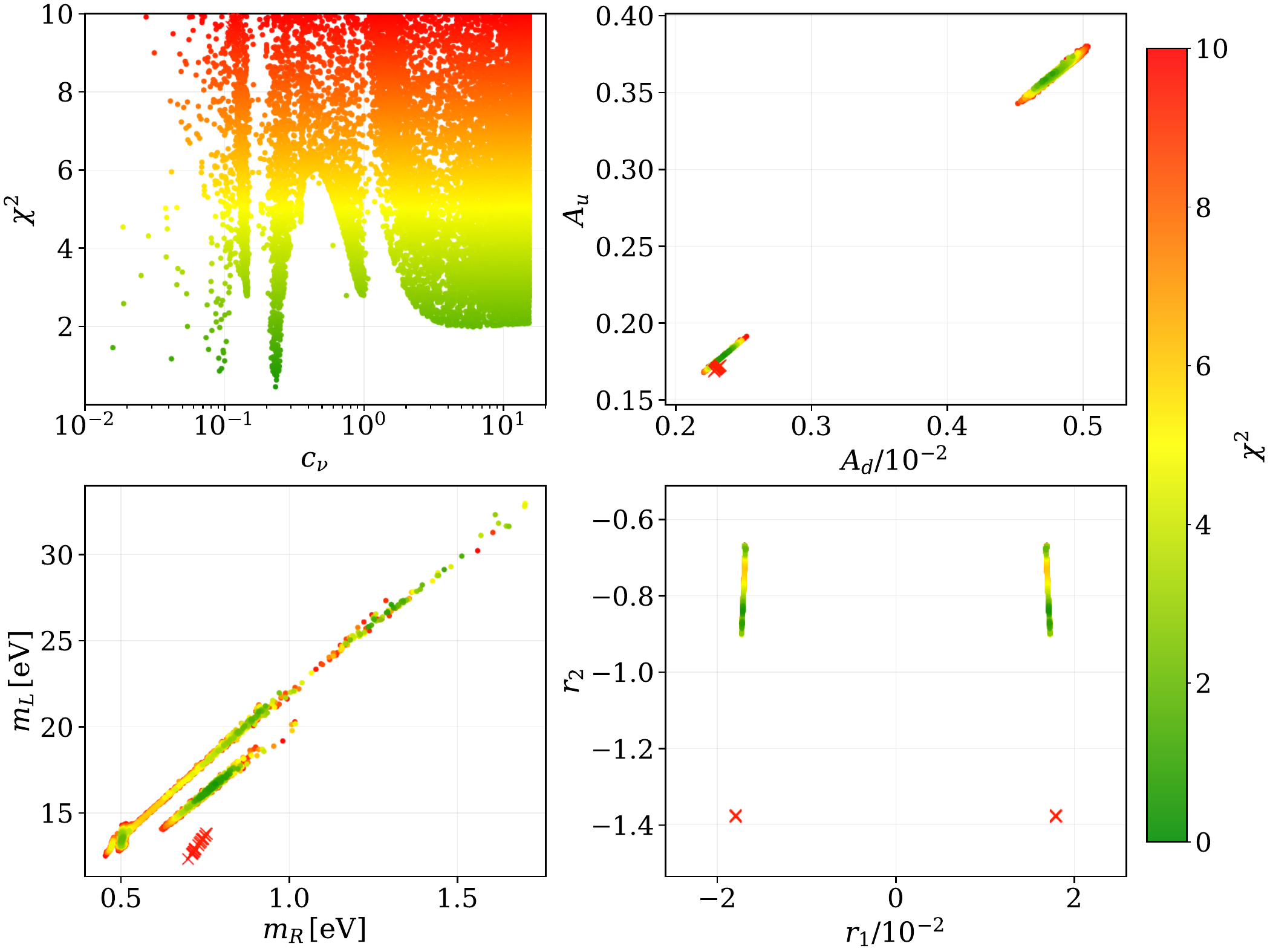}
		\caption{Scan of model parameters with $\chi^2 \leq 10$. Points and crosses denote the scan points with $\eta_d=+1$ and $\eta_d=-1$, respectively.}\label{fig:para}
	\end{center}
	\begin{center}     		
		\includegraphics[width=0.85\textwidth]{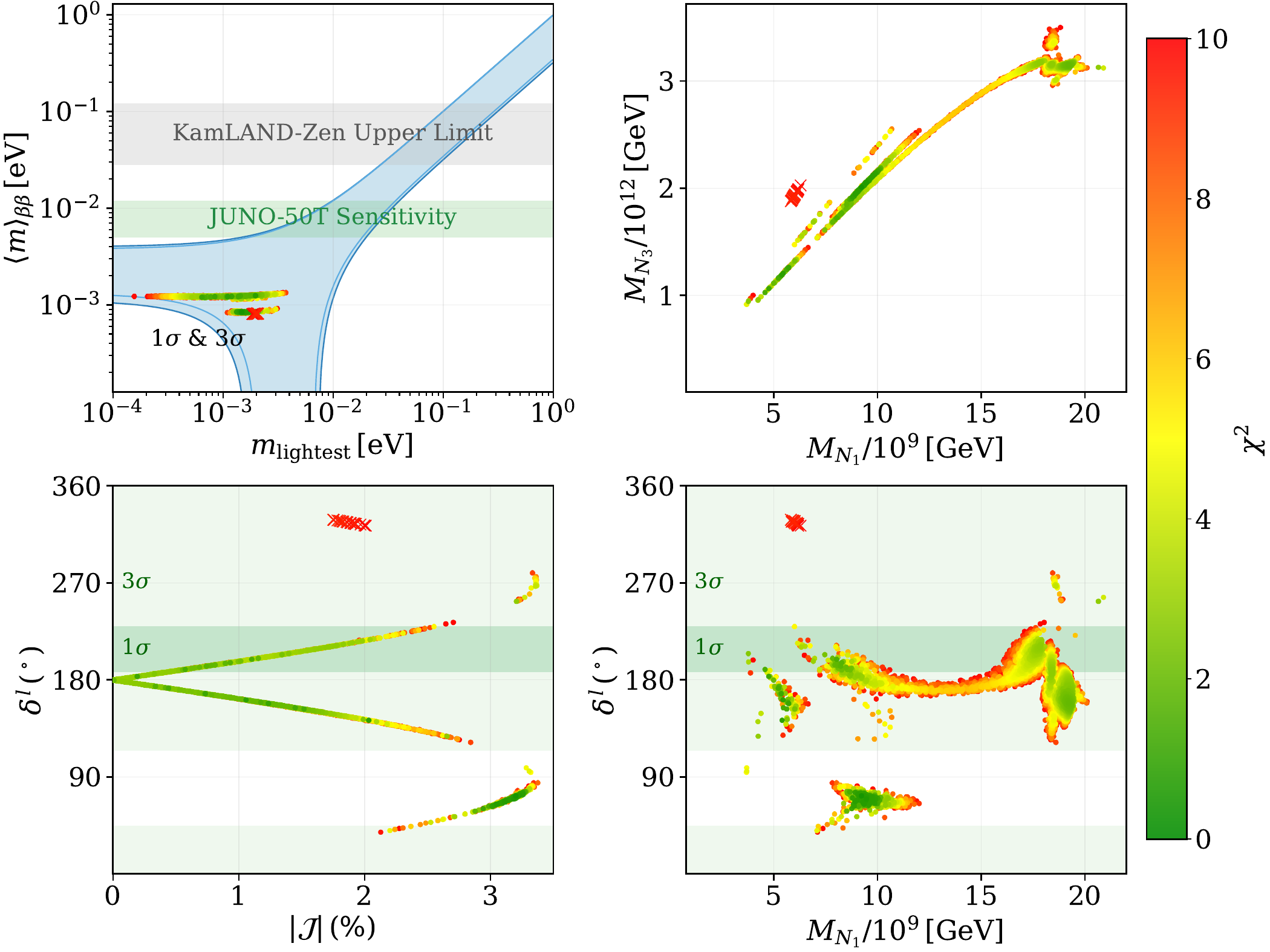}
		\caption{The effective neutrino mass prediction (top left panel) and two-dimensional correlations between predicted observables. Points and crosses denote the scan points with $\eta_d=+1$ and $\eta_d=-1$, respectively. }\label{fig:mee-ml}
	\end{center}
\end{figure}

In summary, we randomly scan seven free model parameters,
$\{A_{u},A_{d},\phi_1,\phi_2, m_L,m_R,c_\nu\}$,
together with the four independent undetermined signs,
$\{\eta_{u},\eta_{d},\eta_{e},\zeta_{d}\}$.
The fit is performed using nine independent observables,
\begin{align}\label{eq:obs}
    {\rm obs}\in \{\theta_{12}^q,\theta_{13}^q,\theta_{23}^q,\delta^q,
    \theta_{12}^l,\theta_{13}^l,\theta_{23}^l,
    \Delta m_{21}^2,\Delta m_{31}^2\},
\end{align}
which contribute to the computation of $\chi^2_{\rm tot}$.

\begin{center}
	\begin{table}[tbp] 
		\renewcommand{\arraystretch}{1.3}	
		\centering
		\begin{tabular}{ >{\centering\arraybackslash}m{11cm}}
			\hline
			 7 free parameters \\
			\hline 
			$\begin{array}{c}
			{\rm Quarks:\qquad} 	A_{u}/y_{t},A_{d}/y_{b}\in(0,1); \  \phi_1,\phi_2\in(0,2\pi), \\
			{\rm Leptons: \qquad} 
			c_\nu\in (10^{-2},15); \ m_L,m_R\in (10^{-1},10^{2}) \ {\rm eV}
			\end{array}$ \\
			\hline
		\end{tabular}
		\caption{Model parameter counting.}
	         \label{tab:Para}	
	\end{table}
\end{center}
Numerical results satisfying $\chi^2_{\rm tot}<10$ are shown in Figs.~\ref{fig:para} and \ref{fig:mee-ml}. Points and crosses denote the scan points with $\eta_d=+1$ and $\eta_d=-1$, respectively. In the present fit, all viable points have $\eta_e=-1$ while no solutions with $\eta_e=+1$ are found.

In the upper panels of Fig.~\ref{fig:para},  the allowed values of $c_\nu$ span a wide range,
$c_\nu\in (0.016,15)$. Most
solutions prefer a large-$c_\nu$ region. The viable points of the quark sector parameters $A_u$ and $A_d$  form two narrow and approximately linear branches, centered around
$\{A_u,A_d\}\sim(0.18,2.3\times10^{-3})$ and
$(0.36,4.7\times10^{-3})$, respectively. 
The linear behavior can be understood from
the parametrizations $A_u=y_t(r+\epsilon)$ and $A_d=y_b(r-\epsilon)$. Since the
CKM constraints require $|\epsilon|\ll |r|$, both $A_u/y_t$ and $A_d/y_b$ are
primarily controlled by the same parameter $r$, leading to the observed linear
correlation. 
In the lower panels of Fig.~\ref{fig:para}, most viable points are found near
$m_L\simeq(12,33)~{\rm eV}$ and
$m_R\simeq(0.45,0.95)~{\rm eV}$. We have not found any viable points with either $m_L$ or $m_R$ vanishing. Thus, both type-I and type-II seesaw are crucial to fit data in UTZT-SO(10).
The parameters $r_1$ and $r_2$ are approximately linearly correlated, compatible with results in Ref.~\cite{Fang:2024qtx}. For the dominant $\eta_d=+1$ branch,
the two signs of $r_1$ correspond to the two assignments
$\zeta_d=\pm1$. In addition, a few $\eta_d=-1$ points appear in a separate
region with $r_2\simeq -1.38$, which represents a new region of parameter space found in this paper. The viable points satisfy
\begin{align}
    A_d \ll A_u,\qquad m_R \ll m_L,\qquad |r_1| \ll |r_2|\,.
\end{align}

The predicted observables are shown in Fig.~\ref{fig:mee-ml}. 
The effective mass of neutrinoless double beta decay is predicted to fall into two narrow bands,
$m_{\beta\beta}\in(0.80,0.93)~{\rm meV}$ and
$(1.14,1.34)~{\rm meV}$,
well beyond the current KamLAND-Zen bound and the future JUNO-50T sensitivity.
The lightest neutrino mass is predicted in the range
$m_{\rm lightest}\in(0.16,3.7)~{\rm meV}$.

We explain why $m_{\beta\beta}$ is confined to narrow bands and is insensitive to $m_{\mathrm{lightest}}$. 
Firstly, we consider the constraints from the two-zero texture of $M_{\nu}$, which is diagonalized by $U_\nu$, 
\begin{equation}
\begin{aligned}
    U_\nu &= U_e \, U_{\mathrm{PMNS}} = P_e\, O_e\, P_{\mathrm{un}}\, V_{\mathrm{PMNS}}\, P_{\mathrm{Maj}} \equiv P_{L\nu}\, V_{\nu}\, P_{R \nu}\,, \label{eq:U_nu}\\
    M_\nu &= U_\nu \,{\rm diag}(m_1, m_2, m_3) \,U_\nu^T \equiv P_{L\nu}\, \overline{M}_\nu\, P_{L\nu} \,,
\end{aligned}
\end{equation}
where $P_{\mathrm{un}} = {\rm diag}(e^{i(\phi_{n1} + \phi_{n3})}, e^{i(\phi_{n2} + \phi_{n3})}, e^{i\phi_{n3}})$ contains the unphysical phases of the PMNS matrix; $P_{\mathrm{Maj}} = {\rm diag}(e^{i\rho}, e^{i\sigma}, 1)$ contains the Majorana phases; the last equality indicates that $U_\nu$ is parametrized in the same way as the PMNS matrix. Due to the non-trivial complex diagonal matrix $P_{\mathrm{un}}$, $P_{L\nu}$ depends on $A_e$, the phases contained in $P_e$, $\phi_{n1}$, $\phi_{n2}$, $\phi_{n3}$, $\delta^l$, and the observables of the lepton sector, while $V_\nu P_{R\nu}$ depends on $A_e$, $\phi_{n1}$, $\phi_{n2}$, $\delta^l$, $\rho$, $\sigma$, and the observables. Given that the Hermitian $Y_\nu$  and the real symmetric $F$ take the two-zero textures, Eq.~\eqref{eq:nu_mass} implies that
\begin{eqnarray}
    M_\nu = \begin{pmatrix}
        0 & M_{12} & 0\\
        M_{12} & M_{22}\,e^{i(\phi_{\nu1} + 2\phi_{\nu2})} & M_{23}\,e^{i\phi_{\nu2}} \\
        0 & M_{23}\,e^{i\phi_{\nu2}} & M_{33} \\
    \end{pmatrix} = P_{L \nu}
    \begin{pmatrix}
        0 & M_{12} & 0\\
        M_{12} & M_{22}\,e^{i\phi_{\nu1}} & M_{23} \\
        0 & M_{23} & M_{33} \\
    \end{pmatrix} P_{L \nu} \,,
\end{eqnarray}
where $M_{12},\,M_{22},\,M_{23}$ and $M_{33}$ are real and $P_{L \nu} = {\rm diag}(e^{-i\phi_{\nu2}}, e^{i\phi_{\nu2}}, 1)$. The two constraints on $P_{L\nu}$ imply that $\phi_{n1}$ and $\phi_{n2}$ can be expressed in terms of $\phi_{n3}$, $\delta^l$, $A_e$, $P_e$, and the observables. The two vanishing elements, $(\overline{M}_\nu)_{11}$ and $(\overline{M}_\nu)_{13}$, provide four constraints on the parameters $m_1$, $\delta^l$, $\rho$, $\sigma$, $\phi_{n1}$, $\phi_{n2}$, $A_e$, and the lepton mixing observables. The parameters $m_1,\, \delta^l,\, \rho,$ and $\sigma$ can be solved in terms of $A_e,\, P_e,\, \phi_{n3}$ and observables. 
Applying these conditions to the expression $m_{\beta\beta} = |m_1 U_{e1}^2 + m_2 U_{e2}^2 + m_3 U_{e3}^2|$, we find that $m_{\beta\beta}$ depends on $A_e$, $P_e$ and the observables in the lepton sector (Note that $\phi_{n3}$ is the overall phase of $U_{\mathrm{PMNS}}$ and thus has no effect on $m_{\beta\beta}$). As shown in Eq.~\eqref{eq:Ye}, $A_e$ and $P_e$ are directly constrained by the quark sector and must be restricted to a narrow range. Compared with $A_e$ and $P_e$, $m_{\mathrm{lightest}}$ is influenced by three additional free parameters from the lepton sector, $\{m_L,\,m_R,\,c_\nu\}$, and thus $m_{\beta\beta}$ is insensitive to $m_{\mathrm{lightest}}$.

The RHN masses are predicted to lie in narrow regions with hierarchy. The lightest one $M_{N_1}\in(0.3,2.0)\times 10^{10}~{\rm GeV}$ and
the heaviest one $M_{N_3}\in(0.9,3.6)\times 10^{12}~{\rm GeV}$. 
These mass scales are potentially compatible with hierarchical thermal leptogenesis, for which successful $N_1$-dominated leptogenesis typically satisfies the Davidson-Ibarra bound, $M_{N_1}\gtrsim10^9~\mathrm{GeV}$ \cite{Davidson:2002qv,Davidson:2008bu}.
The lower panels of Fig.~\ref{fig:mee-ml} show a multi-branch distribution of
$\delta^l$. For the dominant $\eta_d=+1$ solutions, the main branch lies around
$\delta^l\simeq(120^\circ,235^\circ)$ and gives the upper
$m_{\beta\beta}$ band. A smaller branch around
$\delta^l\simeq(40^\circ,85^\circ)$ gives the lower
$m_{\beta\beta}$ band, which is not favored by global-fit data in the $3\sigma$ region.
In addition, a few $\eta_d=-1$ points, absent from
Ref.~\cite{Fang:2024qtx}, appear around
$\delta^l\simeq(320^\circ,330^\circ)$ and
$M_{N_1}\simeq(0.6,0.7)\times 10^{10}~{\rm GeV}$.
However, their $\chi^2$ values are relatively large. These points are shown
only for completeness and should be regarded as marginal solutions.
The Jarlskog invariant can reach the percent level, with
$|\mathcal{J}|\lesssim 3.5\%$. 

As a stability check, we further check the parameter space by imposing a tighter cut 
$\chi^2_{\rm tot}<4$. The effective mass is restricted to two separated
narrow intervals,
$m_{\beta\beta}\in(0.84,0.87)~{\rm meV}$ and
$(1.16,1.29)~{\rm meV}$.
The allowed RHN mass ranges become
$M_{N_1}\in(0.4,\,2.0)\times10^{10}~{\rm GeV}$
and
$M_{N_3}\in(1.0,\,3.2)\times10^{12}~{\rm GeV}$.
The allowed ranges of the CP phase are
$\delta^l\in(54^\circ,80^\circ)$ and
$\delta^l\in(142^\circ,218^\circ)$.
The second interval remains allowed, but fewer points survive the tighter cut.
Compared with the $\chi^2_{\rm tot}<10$ case, the dominant $\eta_d=+1$ regions remain
stable with only moderate narrowing, whereas the marginal solutions associated
with the $\eta_d=-1$ branch disappear.

For comparison, Ref.~\cite{Fang:2024qtx}, using NuFIT 5.3 inputs, found $c_\nu\lesssim1.7$, a connected region $\delta^l\in(90^\circ,230^\circ)$, a single band $m_{\beta\beta}\in(1.15,1.28)~\mathrm{meV}$, 
and a single $\{A_u,A_d\}$ branch around $(0.37,4.8\times10^{-3})$. The present analysis differs from there in both updating the experimental inputs and applying the numerical search strategy. The more comprehensive scan retains the previous $\{A_u,A_d\}$ branch and reveals an additional lower branch around $(0.18,2.3\times10^{-3})$, together with solutions in the $\zeta_d=+1$ branch and a small number of marginal $\eta_d=-1$ solutions. In the resulting viable parameter space, the allowed range extends to $c_\nu\in(0.016,15)$, while $\delta^l$ becomes disconnected into two main intervals, $(40^\circ,85^\circ)$ and $(120^\circ,235^\circ)$. Here, the additional branch at $\delta^l\in(40^\circ,85^\circ)$ was found
in a more extensive parameter scan using the improved numerical algorithm
and is not solely a result of the updated JUNO inputs. Correspondingly, the original $m_{\beta\beta}$ band is retained and slightly broadened to $(1.14,1.34)~\mathrm{meV}$, while an additional lower band emerges at $(0.80,0.93)~\mathrm{meV}$.

\section{UTZT from non-invertible symmetries}\label{sec:4}
Having discussed the phenomenological implications of the UTZT in the previous section, we now turn to its realization. 
Texture zeros realized in the canonical approach via either Abelian or non-Abelian discrete symmetries require more than one generation of Higgs fields \cite{Grimus:2004hf, GonzalezCanales:2012blg}. These  fields introduce a lot of new physical degrees of freedom above the electroweak scale and may greatly affect the RG running to the GUT scale. 
A detailed numerical analysis of the RG running and the consequent prediction of proton decay can be found in \cite{Fang:2024qtx}, where the influence of the repetition number of
Higgs fields ${\bf 10}_H,\overline{\bf 126}_H, {\bf 120}_H$ is discussed in particular. 

Texture zeros realized in NIS's do not introduce any new particles below the GUT scale. 
In this work, we take the economical assumption that only one generation of Higgs fields is introduced, corresponding to the same breaking chain and same particle content as those appearing in the scenario S1) in \cite{Fang:2024qtx} with Higgs repetition number $n_H = 1$. Therefore, the same RG behavior and prediction on the proton decay are expected and we do not repeat it here. We further note that the same breaking chain was earlier discussed in \cite{Bertolini:2009qj,King:2021gmj}.\footnote{The breaking chain in Eq.~\eqref{eq:breaking_chain} is denoted as III in \cite{Bertolini:2009qj} and II3 in \cite{King:2021gmj}, respectively.} Although the particle contents at intermediate scales in these works are not exactly the same, the GUT scales are all predicted to roughly lie in the similar range from $10^{15}$~GeV to $10^{16}$~GeV.
As mentioned in the introduction, the simplest way to realize the UTZT is to impose a non-invertible selection rule of the $Z_3$ gauging of $Z_7$. In this section, we will first discuss the multiplication rules for conjugacy classes in $Z_N \rtimes Z_3$ and then the model building in the $Z_3$ gauging of $Z_7$. We will also introduce the generalized CP symmetry and its implications for the Yukawa couplings. Since NIS is  a type of discrete symmetry, its spontaneous breaking does not generate additional light states and thus  does not affect the RG running.

\subsection{Non-invertible selection rules on $Z_3$ gauging of $Z_N$}
In this subsection, we assume $N$ is prime. Any non-prime $N$ can be expressed as a product of prime factors, i.e., $Z_N \simeq Z_{N_1} \times Z_{N_2} \times \cdots $, which we will comment on at the end of this section. The generators of $Z_N \rtimes Z_3$ are denoted as $a$ and $b$, with the following relations:
\begin{eqnarray}
	a^N = e \,, \quad b^3 = e \,, \quad b^{-1} a b = a^m \,,
\end{eqnarray}
where $e$ is the identity element. Evidently, $Z_3$ is an outer automorphism of $Z_N$ with the mapping $a \to a^m$. The last relation, $ab = ba^m$, implies that
\begin{eqnarray}
	a b^3 = b^3 a^{(m^3)} \quad \Rightarrow \quad a = a^{(m^3)}\,.
\end{eqnarray}
It gives a strong constraint on $m$ and $N$ [\citealp{Dong:2025jra}, \citealp{Qu:2026omn}]
\begin{eqnarray}
	(m^3 - 1) = (m-1)(m^2 + m + 1) = 0 \quad (\text{mod } N) \,.
\end{eqnarray}
Except for the trivial case $m=1$, the above relation is satisfied by $m^2 + m + 1 = 0 \,(\text{mod } N)$. The possible combinations of $N$ and $m$ are
\begin{eqnarray}
	&&N=7 \,, \quad m=2,4 \,, \nonumber\\
	&&N=13 \,, \quad m=3,9 \,...
\end{eqnarray}
Using the relation $ab = ba^m$ and the condition $m^3 \equiv 1 \pmod{N}$, the conjugacy classes of $Z_N \rtimes Z_3$ are given by 
\begin{eqnarray}
	C_1 &=& \{ e \} \,, \nonumber\\
	C^1_3 &=& \{ a,\, a^m = b^{-1} a b ,\, a^{m^2} = b^{-2} a b^2 \} \,, \nonumber\\
	C^k_3 &=& \{ a^k,\, a^{km},\, a^{km^2}\} \,, \nonumber\\
	C^{(1,0)}_3 &=& \{ b,\, ba^{m-1} = a b a^{-1}, ..., ba^{j(m-1)} = a^j b a^{-j},... \} \,, \nonumber\\
	C^{(1,k)}_3 &=& \{ ba^k, ..., ba^{j(m-1)+k},... \} \,, \nonumber\\
	C^{(2,0)}_3 &=& \{ b^2,\, b^2a^{2m-1} = a b^2 a^{-1}, ..., b^2 a^{j(2m-1)} = a^j b^2 a^{-j},... \} \,, \nonumber\\
	C^{(2,k)}_3 &=& \{ b^2a^k, ..., b^2a^{j(2m-1)+k},... \} \,.
\end{eqnarray}
The normal subgroup $Z_N$ generated by $a$ is decomposed into a few conjugacy classes, $C_1$ and $C^k_3$ for $k=1,2,...,\frac{N-1}{3}$, whose fusion rules will be used in the following. The multiplication rules of conjugacy classes are given by
\begin{eqnarray}
	C^k_3 \, C^{l}_3 = C^{l}_3 \, C^k_3 = C^{k + l}_3 + C^{k + ml}_3 + C^{k + m^2l}_3 \,.
\end{eqnarray}

In the following, we will treat NIS's as a bottom-up approach to the texture zeros in the Yukawa coupling. Each field $\phi_k$ is assigned a conjugacy class $C^k_3$ of $Z_N \rtimes Z_3$. A Yukawa coupling $\phi_1 \phi_2 \phi_3$ is allowed if the fusion product of the corresponding conjugacy classes contains the identity class $C_1$, i.e., $C_{\phi_1} C_{\phi_2} C_{\phi_3} \supset C_1$.

For consistency with the kinetic terms, the conjugate field $\phi_k^*$ is labeled by the conjugacy class $[a^{-k}] \equiv C^{N-k}_3$. There is an additional $Z_2$ outer automorphism of the group $Z_N \rtimes Z_3$ with the mapping $a \to a^{-1}$, corresponding to the charge conjugation. Its transformation on the conjugacy classes of a group $G$ is defined as 
\begin{eqnarray}
	[g_i] \to X_{ij}^{\text{CP}}\,[g_j^{-1}] \,,
\end{eqnarray}
where $X^{\text{CP}}$ is a unitary matrix. In the case of the $Z_2$ gauging, $a$ and $a^{-1}$ belong to the same conjugacy class, making this a trivial case with respect to the generalized CP symmetry. Combining the parity transformation with the charge conjugation, we can define the generalized CP transformation on a field as \cite{Kobayashi:2025wty}
\begin{eqnarray}
	\phi(x) \to X^{\text{CP}}\,\phi(x_{\rm P})^* \,,
\end{eqnarray}
with $x = (t,x)$ and $x_{\rm P} = (t,-x)$. If there is a global discrete symmetry $G_f$, an additional consistency condition should be satisfied. The generalized CP transformation reduces to the canonical CP transformation discussed in section~\ref{sec:2}, if $X^{\text{CP}}=I$.

We conclude this subsection with an example of the $Z_3$ gauging of $Z_7$ with $m=2$. The conjugacy classes are
\begin{equation}
	\begin{aligned}
	C_1 =&\, \{ e \} \,, \nonumber\\
	C^1_3 =&\, \{ a,\, a^2,\, a^4 \} \,, \nonumber\\
	C^2_3 =&\, \{ a^3,\, a^6,\, a^5\} \,,
	\end{aligned}
\end{equation}
whose multiplication rules are listed in Table~\ref{tab:Z7}. Under the generalized CP transformation, conjugacy classes transform as $C_1 \leftrightarrow C_1$ and $C^1_3 \leftrightarrow C^2_3$.
\begin{table}[h]
    \centering
	\renewcommand{\arraystretch}{1.3}
    \begin{tabular}{|c||c|c|c|}
    \hline
          & $C_1$ &  $C_3^1$ & $C_3^2$  \\
        \hline
		\hline
		$C_1$ & $C_1$ & $C_3^1$ & $C_3^2$  \\
		\hline
        $C_3^1$ & $C_3^1$ & $C_3^1 + 2C_3^2$ & $3C_1 + C_3^1 + C_3^2$  \\
		\hline
        $C_3^2$ & $C_3^2$ & $3C_1 + C_3^1 + C_3^2$ & $2C_3^1 + C_3^2$  \\
        \hline
    \end{tabular}
    \caption{Multiplication rules for the conjugacy classes in the $Z_3$ gauging of $Z_7$.}
    \label{tab:Z7}
\end{table}

\subsection{Model building under non-invertible selection rules}
\begin{table}[ht] 
	\centering
	\begin{tabular}{ m{2cm}<{\centering} m{4cm}<{\centering} m{4cm}<{\centering}}
		\hline \hline \\[-4mm]
		 & Particles in $SO(10)$ & Class assignment in $Z_3$ gauged $Z_7$ \\[-4mm]\\\hline \\[-4mm]
		Fermions & ${\bf 16}_F^i$ & $(C_1,C_3^1,C_3^2)$ \\[-4mm]\\\hline 
		\\[-3mm]	
		& ${\bf 10}_H,{\bf 120}_H$ & $C_3^2$ \\[-3mm] \\\cline{2-3} \\[-4mm]
		Higgses & $\overline{\bf 126}_H$ & $C_3^2$
		\\[-5mm] \\\cline{2-3} \\[-4mm]
		& ${\bf 54}_H$ & $C_3^2$ \\[-4mm] \\\cline{2-3} \\[-4mm]
		& ${\bf 45}_H$ & $C_3^2$ \\[-4mm] \\\hline
		\hline
	\end{tabular}
	\caption{Particle content of $SO(10)$ with the $Z_3$ gauging of $Z_7$. Here, the superscript $i$ labels three fermion generations, $i=1,2,3$.}
	\label{tab:particle_charge_assignment}
\end{table}

To realize the UTZT, we assign different conjugacy classes of $Z_7 \rtimes Z_3$ to the Higgs fields and three generations of fermions, as shown in Table~\ref{tab:particle_charge_assignment}. Given the fusion rules in Table~\ref{tab:Z7}, it is straightforward to see that the three Yukawa matrices in Eqs.~\eqref{eq:Yukawa_couplings_SO(10)} share the same two-zero texture. For example, the $(1,3)$ element of $A$ is zero because the fusion product of the corresponding conjugacy classes, $C_1\, C_3^2\, C_3^2$, does not contain the identity class $C_1$. The same argument applies to the other elements in the Yukawa matrices. In the Yukawa sector, the class assignment for fermions and for ${\bf 10}_H,{\bf 120}_H, \overline{\bf 126}_H$ in Table~\ref{tab:particle_charge_assignment} is unique up to the trivial interchange of $C_3^1$ and $C_3^2$.

For completeness, we have also considered an example of the class assignments for the Higgs fields ${\bf 54}_H$ and ${\bf 45}_H$, which are responsible for the spontaneous breaking of $SO(10)$ to $G_{422}^{\rm C}$ and $G_{422}^{\rm C}$ to $G_{3221}$, respectively. Table~\ref{tab:Z7} implies that arbitrary couplings involving three or more Higgs fields are not forbidden by the selection rules, provided all fields are assigned either $C_3^1$ or $C_3^2$. Quadratic terms formed by fields with their conjugation, such as $({\bf 54}_H {\bf 54}^*_H)$, $({\bf 45}_H {\bf 45}^*_H)$ and $({\overline{\bf 126}}_H \overline{\bf 126}_H^*)$, are allowed regardless of class assignments. The class assignment for ${\bf 54}_H$ and ${\bf 45}_H$ are not unique, but this does not affect the UTZT in flavor space, and we therefore do not discuss them in detail.

The same two-zero texture of the neutrino mass matrix was realized in an earlier work~\cite{Qu:2026omn}, where the authors assumed that neutrino masses are generated by either the Weinberg operator or type-I seesaw mechanism without considering the origin of $B-L$ breaking. Since no charge can be assigned to coefficients of the Weinberg operator or Majorana mass parameters, they concluded that the $Z_3$ gauging of $Z_{13}$ is the minimal symmetry that realizes the same two-zero texture, which differs from our hypotheses and results. As an example, consider the realization of texture zeros in the RHN mass term, $\frac{1}{2}(M_N)_{\alpha \beta}\overline{\nu^C_{R \alpha}} \nu_{R \beta}$. In our model, $M_N$ arises from the VEV of a Higgs that carries both the $B-L$ and NIS charges. After this Higgs acquires a VEV, both the $B-L$ symmetry and the NIS are spontaneously broken. By contrast, the mass matrix $M_N$ in Ref.~\cite{Qu:2026omn} carries no NIS charge and breaks $B-L$ explicitly; consequently, the NIS remains unbroken in their construction.

Finally, we briefly discuss alternative ways to realize the UTZT. In the case of one generation of Higgs fields, it is impossible to realize the UTZT with either $Z_N \rtimes Z_2$ (a brief notation for the $Z_2$ gauging of $Z_N$), $Z_{13} \rtimes Z_3$, $(Z_2 \times Z_2^\prime \times Z_2^{\prime\prime}) \rtimes Z_3$ or $(Z_3 \times Z_3^\prime) \rtimes Z_3$. However, the $Z_2$ gauging of $Z_N$ ($N = 6, 7, 8, \dots$) can realize the UTZT if two generations of Higgs fields are included. For example, in the case of the $Z_2$ gauging of $Z_6$, one option is to label the first generation of Higgs fields with $[a^2]$, the second with $[a^3]$, and the three generations of fermions with $[e], [a^2], [a]$, where $a$ is the generator of $Z_6$ and $e$ is the identity element. We will not discuss the details of these alternative realizations due to the enormous freedom introduced by the additional Higgs generation.

\section{Conclusions}\label{sec:5}

Texture zeros remain one of the most enduring approaches to the flavor puzzle in particle physics. A particularly predictive pattern is the universal two-zero texture (UTZT), in which all fermion mass matrices possess two vanishing $(1,1)$ and $(1,3)$ entries. The UTZT is providing a compelling description of fermion masses and mixing due to its remarkable consistency with current experimental data. Embedding it into SO(10) GUT further enhances its predictive power by exploiting the quark-lepton correlation of GUT. In this work, we presented a state-of-the-art analysis of the UTZT-SO(10). We took into account the new data released from the JUNO reactor neutrino experiment where the normal ordering of light neutrino masses are indicated in $2.3\sigma$. We showed that the UTZT-SO(10) fits very well with the precision data.
In addition, we investigated the possible origin of the UTZT from non-invertible symmetries and discussed explicit symmetry realizations of the texture structure. 

Fixing the Yukawa eigenvalues of six quarks and three charged leptons at their best-fit values reduces the number of independent free parameters to seven. These parameters are then used to fit nine observables, namely the four CKM mixing parameters, the three leptonic mixing angles, and the two neutrino mass-squared differences. Our numerical analysis demonstrates that the UTZT within the SO(10) framework provides an excellent description of the latest experimental data. The Dirac CP-violating phase regarded as a prediction of the fitting is restricted mainly in two regimes, $\delta^l \simeq (120^\circ, 235^\circ)$ and $(40^\circ, 85^\circ)$. The former region is in very good agreement with current global-fit results, whereas the latter is disfavored at the $3\sigma$ level. 
The effective Majorana mass governing neutrinoless double-beta decay is predicted in very narrow bands, $m_{\beta\beta}\simeq (0.80,0.93)$~meV and $(1.14,1.34)$~meV, associated with the two disconnected $\delta^l$ regions. 
The RHN mass spectrum is also strongly constrained. The lightest one is roughly around $10^9$~GeV and the heaviest one is of order $10^{12}$~GeV.

Although texture zeros can be generated through conventional Abelian or non-Abelian flavor symmetries, realizing them via non-invertible symmetries offers a unique advantage: no additional low-energy degrees of freedom are required.
We demonstrated the UTZT via non-invertible selection rules on the $Z_3$ gauging of $Z_N$. In particular, we presented explicit class assignment based on the $Z_3$ gauging of $Z_7$, which, so far, is the minimal symmetry to generate this type of two-zero texture. We further comment that the UTZT can also be realized in the $Z_2$ gauging of $Z_N$ if more than one generation of Higgses are included in the model.

\section*{Acknowledgement}

This work was supported by National Natural Science Foundation of China (NSFC) under Grant No.~12535007,  No.~12547104, and Zhejiang Provincial Natural Science Foundation of China under Grant No.~LDQ24A050002.

\end{document}